# Monte Carlo Investigation of Anomalous Transport in presence of a Discontinuity and of an Advection Field

M. Marseguerra [1], A. Zoia[2]

*Department of Nuclear Engineering, Polytechnic of Milan, Via Ponzio 34/3, 20133 Milan, Italy*

## Abstract

Anomalous diffusion has recently turned out to be almost ubiquitous in transport problems. When the physical properties of the medium where the transport process takes place are stationary and constant at each spatial location, anomalous transport has been successfully analyzed within the Continuous Time Random Walk (CTRW) model. In this paper, within a Monte Carlo approach to CTRW, we focus on the particle transport through two regions characterized by different physical properties, in presence of an external driving action constituted by an additional advective field, modeled within both the Galilei invariant and Galilei variant schemes. Particular attention is paid to the interplay between the distributions of space and time across the discontinuity. The resident concentration and the flux of the particles are finally evaluated and it is shown that at the interface between the two regions the flux is continuous as required by mass conservation, while the concentration may reveal a neat discontinuity. This result could open the route to the Monte Carlo investigation of the effectiveness of a physical discontinuity acting as a filter on particle concentration.

## 1. Introduction

A general approach to the analysis of transport phenomena is based on Continuous-Time Random Walk (CTRW) [1-6], in which the travel of a particle (a walker) in a medium is modelled as a series of jumps of random lengths, separated by random waiting times. The theory of CTRW with algebraically decaying probability distribution functions (pdf's) has been originally introduced in Physics in a series of seminal papers by Weiss, Scher, Montroll and co-workers [4,7-8] in the late 1960s to explain evidences of anomalous diffusion occurring in the drift-diffusion processes in amorphous semiconductors. The diffusion is called anomalous if the mean squared displacement (MSD) is not linearly proportional to time *t* as in the standard Fickian case, but to powers of *t* larger (superdiffusion) or smaller (subdiffusion) than unity. More recently, anomalous diffusion has turned out to be quite ubiquitous in almost every field of science (see e.g. [1] and [2] for a detailed review) and the CTRW model has been applied with success to interpret the experimental results and to make predictions on the evolution of the examined systems [3-5,8]. Such applications concern among the others e.g. the behaviour of chaotic Hamiltonian systems (with application to the transport of charged particles in turbulent plasma) [9-10], the evolution of financial markets [11], the dynamics of ad-atoms on the surface of a solid [12] or the underground transport of contaminant particles in presence of rock fractures and porosity [13]. Much attention has been paid to this last topic, as many laboratory-scale experiments as well as direct field measurements have evidenced subdiffusive and superdiffusive behaviours of migrating particles.

The present paper focuses on the Monte Carlo approach to the subdiffusive transport of particles in a medium constituted by two regions characterized by different physical properties, in presence of an additional (constant) velocity field. In subdiffusive transport [13-16,1-3], a finite-variance distribution – e.g. a Gaussian pdf – and an algebraically decaying distribution with infinite mean (instead of the traditional exponential pdf, which would lead to Fickian diffusion) are assumed for the jump lengths and for the flight times of the particles, respectively. Thus, the particles will have a non-vanishing probability of extremely long times between successive visited locations and this may be physically interpreted as particles were trapped in the medium for long times. A large number of "walkers" are followed along their simulated travels through the medium: their ensemble behaviour yields the flux $\phi(x,t)$ and the resident concentration $P(x,t)$. The core of the present analysis is the investigation of the joint effect of the discontinuity, which introduces a macroscopic heterogeneity in the traversed medium, and of the advective field, which acts as an external driving action. The discontinuity has the effect of modifying the distributions of the jump lengths and of the flight times which rule the walker's travels whenever a particle crosses the discontinuity itself. The advective field is represented in form of an external forward velocity, framed within either the Galilei invariant or Galilei variant approaches. This velocity has the effect of encouraging the particles to move forwardly in the velocity direction: nonetheless, this advection contribution is in competition with the coexisting diffusion process, which might temporarily drive a particle backward, against the bias. Moreover, while in the standard CTRW model the walker is assumed to be trapped for the whole waiting time at the starting point of each jump and then be suddenly transferred to the new sojourn location with infinite speed, we will assume a uniform linear motion between successive interactions.

---

[1] *Corresponding Author. Tel: +39 02 2399 6355 . Fax: +39 02 2399 6309*
*Email address: marzio.marseguerra@polimi.it (Marzio Marseguerra)*

[2] *Email address: andrea.zoia@polimi.it (Andrea Zoia)*

The paper is structured as follows: in Section 2 we introduce the statement of the problem, i.e. anomalous transport across a boundary. In Section 3 the method adopted in sampling jumps across the discontinuity is illustrated through an example involving memoryless Markovian particles such as neutrons. Then, in Section 4 this method is extended to semi-Markovian processes and applied to subdiffusive transport in presence of advection, in both Galilei invariant and Galilei variant schemes. Section 5 is devoted to a brief presentation of the Monte Carlo collection method which allows to evaluate particle resident concentration and flux. In Section 6 we illustrate the effects of the discontinuity by means of some examples of Monte Carlo simulations, showing that in the general case, when the travel distributions parameters are freely chosen, the interface introduces a neat discontinuity in the resident concentration. Conclusions are finally drawn in Section 7 and some considerations on the advection field are resumed in Appendix.

**2. Statement of the problem: walking across a discontinuity**

Anomalous transport has been successfully described in the framework of the CTRW approach, in which a walker diffuses by performing a sequence of semi-Markovian jumps in space and time. For sake of simplicity, the transport is usually thought to occur on a one-dimensional space. We add to the purely diffusive contribution a constant velocity field and further assume that the transport takes place in a medium constituted by two different layers, say $L_1$ on the left and $L_2$ on the right, the discontinuity occurring at $x=x_d$. Each layer is characterized by different physical properties, i.e. by different parameters of the jump lengths and/or waiting times distributions.

The semi-Markovianity assumption amounts to saying that, along the walker's trajectory, each jump starts without memory of the way in which the jumping point has been reached. The process is regenerated at each visited point of the trajectory and it is then ruled by space and time probability density functions (pdf's) which in general depend on the time and space intervals from the jump origin and on the features of the surrounding medium.

In each layer $i$ ($i=1,2$) the walker undergoes a succession of jumps, each performed in a time interval $\Delta t$ drawn from a pdf $w_i(\Delta t)$: the jump lengths are given by the sum of a diffusive stochastic contribution $\Delta x$ independently drawn from a pdf $\lambda_i(\Delta x)$ (with zero mean) and of an advective contribution due to the velocity field: advection is usually modelled within the so-called Galilei invariant or Galilei variant schemes [1]. In the first case, a contribution $v\Delta t$, where $v$ is the constant velocity field, is added to the space variable $\Delta x$ sampled from $\lambda_i(\Delta x)$, which represents diffusion. In the latter, at each jump a constant bias $\mu$ is added to $\Delta x$. The physical meaning of the field $v$ and of the bias $\mu$ and the applicability of the two approaches is discussed in detail e.g. in [17-19,1] and some considerations will be presented in the Appendix. Both these schemes admit formal analytical solutions in the framework of the Fractional Advection-Diffusion Equation (FADE) model [1], which can be derived as an approximation of the CTRW approach. In particular, an intensive work has been performed by Scher, Berkowitz, Margolin and co-workers in the context of the Galilei variant scheme in order to improve the accuracy of the solutions of the CTRW model and to interpret experimental evidences, with application to the problem of contaminant transport in porous and fractured media [13-20,3].

As long as a jump is entirely performed within the same layer (i.e. when the discontinuity plays no role in the transport), the occurrence of a velocity field does not dramatically change the complexity of the CTRW model. The situation is quite different when the particles trajectories start in a layer and end in the other one, thus passing through the discontinuity. Here the situation is similar to that of the time paradox described by Feller [21], who put in evidence that, even in the trivial case of a succession of exponential time intervals with mean value $\tau$, the distribution of those intervals which went across a fixed time point was "different", i.e. no more exponential and with a mean value $2\tau$. Similarly, in our transport problem, it turns out that the space and time distributions of the jumps across the heterogeneity are different and may even cause an important discontinuity at $x_d$ in the shape of the resident concentration distribution $P(x,t)$. Moreover, the present analysis stands out from the usual one in the CTRW context [1-2,4-7], in which it is assumed that the walker, after a spatial jump, rests in the reached location for the time interval $\Delta t$ and then performs another instantaneous jump (with infinite speed) to the successive location. Instead, we will assume that at each jump the walker moves linearly with constant velocity – given by the ratio of the jump length to the flight time – between successive interactions: under this assumption, we will have a spectrum of velocities as a result of the stochastic variability of both the jump lengths and the flight times. To the authors' best knowledge, no fractional advection-diffusion scheme exists for such a situation, so that the Monte Carlo simulation seems a good method to investigate the matter.

## 3. The neutron transport: a Markovian case

In order to make clear how we will sample a random jump across an heterogeneity, we consider at first this problem in the well-known exponential case, with reference to the monoenergetic neutron flights. A neutron trajectory consists of straight flights between successive (random) collisions with nuclei. Each flight length is sampled from an exponential cumulative distribution function (cdf) $W(x/\sigma)$, whose parameter $\sigma$ [cm$^{-1}$] is the inverse of the mean free path, which we assume to be dependent only on the density of the nuclei. The pdf of the flight length being exponential, the stochastic process is ensured to be Markovian. Assume that layer $L_1$ is characterized by the parameter $\sigma_1$. Starting from $x_0=0$ in $L_1$, a neutron flight of length $x^*$ is sampled from $W(x|\sigma_1)$ by the inverse transform technique, utilizing a random $R_1$ uniform in $[0,1)$, so that $x^*$ is solution of

$$R_1 = W(x^* | \sigma_1) = 1 - e^{-\int_0^{x^*} \sigma_1(u)du} = 1 - e^{-\sigma_1 x^*} \tag{1}$$

Suppose now that, after having travelled a chord of length $c<x^*$, the neutron enters layer $L_2$, which has a different density and is then characterized by a different parameter $\sigma_2$. The old value $x^*$ is discarded and the length of the flight varies from $x^*$ to a value $x_1>c$, to be determined. There are two possible ways of determining $x_1$, according to whether we choose to determining the remaining portion of the path in $L_2$ by utilizing the same $R_1$ already sampled or by sampling a new uniform random $R_2$. In the first case, the cdf pertaining to the two pieces of the jump is

$$W(x | c, \sigma_1, \sigma_2) = 1 - e^{-\int_0^x \sigma(u)du} = 1 - e^{-[\sigma_1 c + \sigma_2(x-c)]} \tag{2}$$

Equating this cdf to $R_1$ yields

$$x_1 = \frac{(\sigma_2 - \sigma_1)c - \ln(1 - R_1)}{\sigma_2} \tag{3}$$

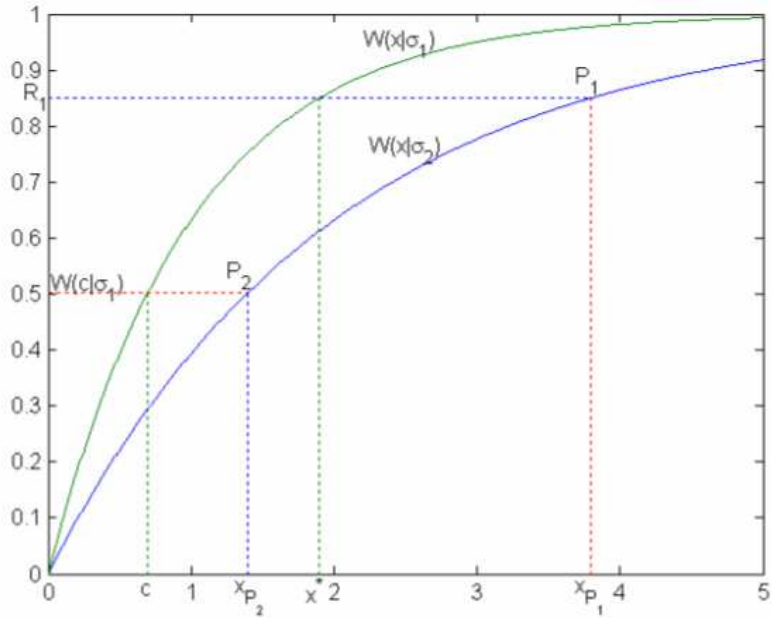

*Figure 1. Sampling a jump from two different exponential distributions.*

The same result may be attained by tackling the problem from a different point of view, the one which we shall adopt for the case of semi-Markovian space and time distributions.

Figure 1 shows the two exponential cdf's together with the $R_1$ value and the corresponding jump length $x^* = W^{-1}(R_1 | \sigma_1) > c$. The unused fraction of $R_1$, from $W(c|\sigma_1)$ to $R_1$, must be utilized by $W(x|\sigma_2)$, on which it gives rise to the points $P_1$ and $P_2$, whose abscissas difference yields the rest of the jump, *viz.*,

$$W(x_{P1} | \sigma_2) = R_1 \quad ; \quad W(x_{P2} | \sigma_2) = W(c | \sigma_1) \tag{4}$$

from which

$$\sigma_2 x_{P1} = -\ln(1-R_1) \; ; \qquad \sigma_2 x_{P2} = \sigma_1 c$$

and finally

$$x_1 = c + (x_{P1} - x_{P2}) \qquad (5)$$

coincident with Eq. (3).

As mentioned above, the portion of the path in $L_2$ may be alternatively determined with a new random number. In the present case, the Markovian, memoryless feature of the process allows one to sample this second portion from $W(x/\sigma_2)$ with a new random $R_2$, uniform in $[0,1)$, independently of the already travelled distance $c$. By adding the two contributions, we get

$$x_1' = c - \frac{1}{\sigma_2}\ln(1-R_2) \qquad (6)$$

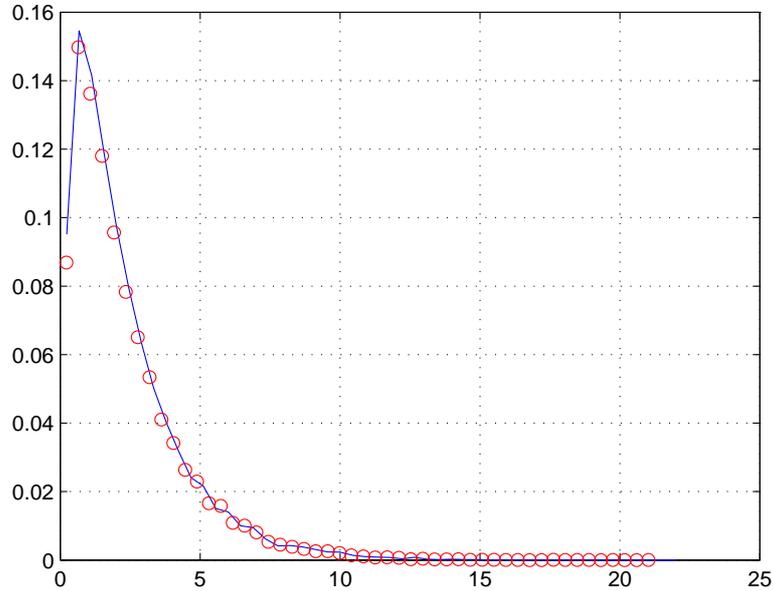

Figure 2. Distributions of $x_1$ (circles) and $x'_1$ (solid line)

Clearly, the two random variables $x_1$ and $x_1'$ so obtained are equal in distribution. An example of the two distributions is shown in Figure 2. It appears that the exponential feature pertaining to the jump lengths in an homogeneous layer is lost.

## 4. Subdiffusive transport: a semi-Markovian case

We begin by tackling the case of the Galilei invariant scheme: at each jump, the effect of the advection is modelled by adding to the diffusion component $\Delta x$ the term $v\Delta t$, where $v$ is the velocity due to the external field and $\Delta t$ is the jump time length. The situation is rather involved, as the advection contribution to each jump is stochastic and it is also possible that a particle gets stuck at the discontinuity. This latter event may occur – as we shall see – when the competition between an event of backward diffusion and the forward advection is such that the particle, once reached the discontinuity, cannot move forward nor backward.

The diffusion contribution to each jump length in layer $L_i$ is assumed to be sampled from any finite-variance distribution. For sake of simplicity, here we adopt a Gaussian $(0, \Sigma_i^2)$ with pdf and cdf $\lambda(\Delta x/\Sigma_i)$ and $\Lambda(\Delta x/\Sigma_i)$, respectively. To this contribution we must add the one of the advection. To simplify the notations, in the following we shall write $x$ and $t$ instead of $\Delta x$ and $\Delta t$, viz.,

$$\lambda(x\,|\,\Sigma_i) = \frac{1}{\sqrt{2\pi}\Sigma_i} e^{-\frac{1}{2}\frac{x^2}{\Sigma_i^2}} \quad ; \quad \Lambda(x\,|\,\Sigma_i) = \int_{-\infty}^{x} \lambda(z\,|\,\Sigma_i)dz \tag{7}$$

As for the distribution of the flight times, the Central Limit Theorem ensures that any finite mean pdf gives rise to Fickian diffusion. A way to obtain a subdiffusive behaviour, yet preserving stability, consists in sampling the times from an infinite mean distribution. In this respect, a widely adopted choice (which grants the applicability of the Lévy-Gnedenko Theorem and thus ensures the stability of the resulting $P(x,t)$ [21-26,1]) is a power law distribution, characterized by fat tails, with pdf and cdf $w(t|\tau_i,\alpha_i)$ and $W(t|\tau_i,\alpha_i)$, respectively. Here we assume that for $0 \leq t \leq \tau_i$ we have

$$w(t\,|\,\tau_i,\alpha_i) = p_i \frac{2}{\tau_i^2} t \quad ; \quad W(t\,|\,\tau_i,\alpha_i) = p_i \left(\frac{t}{\tau_i}\right)^2 \tag{8}$$

and, for $\tau_i \leq t < \infty$

$$w(t\,|\,\tau_i,\alpha_i) = (1-p_i)\frac{\alpha_i}{\tau_i}\left(\frac{\tau_i}{t}\right)^{\alpha_i} ; \; W(t\,|\,\tau_i,\alpha_i) = p_i + (1-p_i)\int_{\tau_i}^{t} w(z\,|\,\tau_i,\alpha_i)dz = 1 - (1-p_i)\left(\frac{\tau}{t}\right)^{\alpha_i} \tag{8'}$$

where $p_i = W(\tau_i\,|\,\tau_i,\alpha_i) = \frac{\alpha_i}{(2+\alpha_i)}$ is the probability of a waiting time less or equal to $\tau_i$. To summarize, then, the two layers $L_i$ may be distinguished by different $\alpha_i$, $\sigma_i$ and $\tau_i$.

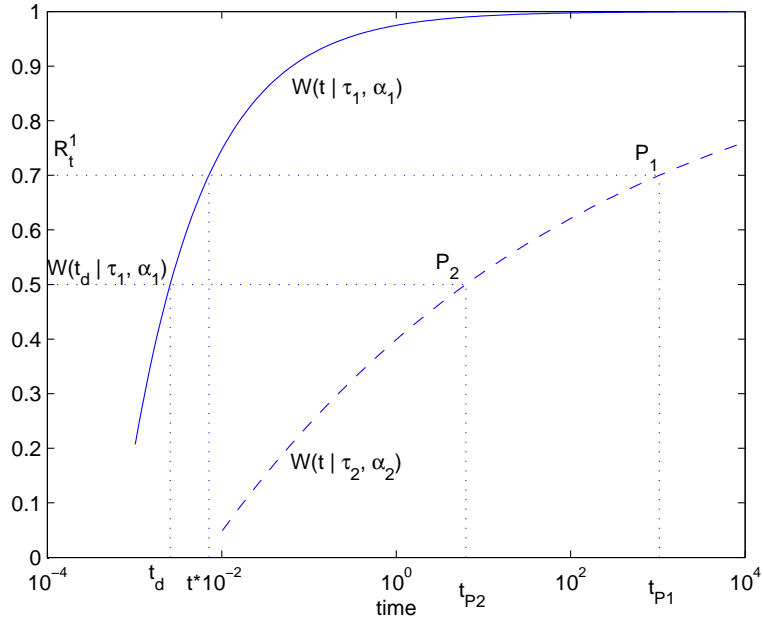

*Figure 3. Power law cdf of jump durations. Solid line: layer $L_1$, $\alpha_1=0.5$; dashed line: layer $L_2$, $\alpha_2=0.1$*

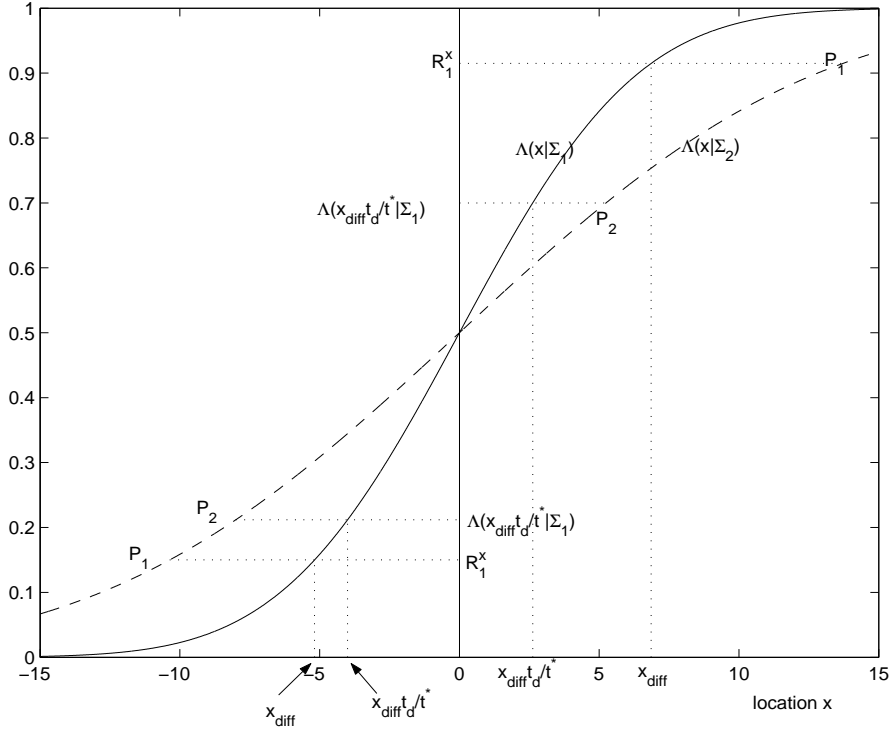

*Figure 4. Gaussian cdf of jump lengths. Solid line: layer $L_1$, $\Sigma_1=5$; dashed line: layer $L_2$, $\Sigma_2=10$*

Let us now consider a walker's jump starting at $x_0 \in L_1$ at time $t_0$. The jump is totally determined by sampling two random numbers $R_1^x$ and $R_1^t$, both uniform in *[0,1)*. The duration of the jump and the diffusion contribution to the jump length are $t^* = W^{-1}(R_1^t \mid \tau_1, \alpha_1)$ and $x_{diff} = \Lambda^{-1}(R_1^x \mid \Sigma_1)$, respectively. Taking into account the advection contribution, the total jump length would be $x_{diff} + vt^*$ if the new location were still in $L_1$. On the contrary, suppose that the walker crosses the discontinuity located at $x_d$ after having utilized the portions $t_d = \frac{x_d - x_0}{x_{diff} + vt^*} t^*$ and $x_{diff} \frac{t_d}{t^*}$ of the assigned $t^*$ and $x_{diff}$ and the portions $W(t_d \mid \tau_1, \alpha_1)$ and $\Lambda(x_{diff} t_d / t^* \mid \Sigma_1)$ of the assigned $R_1^x$ and $R_1^t$. Going along the line given in the simple exponential case, the rest of the jump must be then effectuated by completing the utilization of $R_1^x$ and $R_1^t$ according to the space and time cdf's pertaining to $L_2$.

First, consider the time duration. With reference to Figure 3, the time duration to be added to $t_d$ is $t_{P_1} = W^{-1}(R_1^t \mid \tau_2, \alpha_2)$, the travel time if totally effectuated in $L_2$, decreased by $t_{P_2}$, solution of $W(t_{P_2} \mid \tau_2, \alpha_2) = W(t_d \mid \tau_1, \alpha_1)$, which is the time corresponding to the amount of $R_1^t$ not utilized in $L_1$. Then, the total jump duration is

$$t_1 = t_d + \left[ W^{-1}\left(R_1^t \mid \tau_2, \alpha_2\right) - W^{-1}\left(W(t_d \mid \tau_1, \alpha_1) \mid \tau_2, \alpha_2\right) \right] \qquad (9)$$

Note that, in absence of discontinuity, i.e. for $\tau_1 = \tau_2 = \tau$ and $\alpha_1 = \alpha_2 = \alpha$, we have $W^{-1}\left(W(t_d \mid \tau, \alpha) \mid \tau, \alpha\right) = t_d$ and the expression for $t_1$ becomes the standard one for a jump totally evolving in an homogeneous medium, i.e. $t_1 = W^{-1}(R_1^t \mid \tau, \alpha)$.

Next, consider the space travelled after the start of the jump (cf. Fig.4). The situation is rather involved, because of the interplay between the contributions to the jump length due to the diffusion – half of which are negative – and those due to the advection. In the specific jumps we are considering, namely those which go across the discontinuity, the length travelled in $L_1$ during $t_d$ is $c = x_d - x_0$, sum of the diffusion component $x_{diff} t_d / t^*$ and of the advection component $vt_d$. Correspondingly, the portion of $R_1^x$ spent in $L_1$ is $\Lambda(x_{diff} t_d / t^* \mid \Sigma_1)$ and the remaining interval, from $\Lambda(x_{diff} t_d / t^* | \Sigma_1)$ to $R_1^x$, must be utilized in connection with $\Lambda(x | \Sigma_2)$ to determine the additional diffusive component travelled in $L_2$. To this aim, we must further distinguish whether $x_{diff}$ is positive or negative.

If $x_{diff}$ is positive $(R_1^x > 0.5)$ (cf. Fig.4, top), then $0 < x_{diff} t_d / t^* < x_{diff}$ and correspondingly $0.5 \leq \Lambda(x_{diff} t_d / t^* | \Sigma_1) < R_1^x$. The additional diffusive component is directed as the advection velocity and it is given by the difference between the abscissas of the points P$_1$ and P$_2$ on $\Lambda(x | \Sigma_2)$. To this quantity we must add the contribution $v(t_1 - t_d)$ of the advection, so that the jump ends at

$$x_1 = \{x_0 + x_{diff} t_d / t^* + vt_d\} + \{[\Lambda^{-1}(R_1^x | \Sigma_2) - \Lambda^{-1}(\Lambda(x_{diff} t_d / t^* | \Sigma_1) | \Sigma_2)] + v(t_1 - t_d)\}$$
$$= x_d + \{[\Lambda^{-1}(R_1^x | \Sigma_2) - \Lambda^{-1}(\Lambda(x_{diff} t_d / t^* | \Sigma_1) | \Sigma_2)] + v(t_1 - t_d)\} \tag{10}$$

In absence of discontinuity, i.e. for $\Sigma_1 = \Sigma_2 = \Sigma$, we have $\Lambda^{-1}(\Lambda(x_{diff} t_d / t^* | \Sigma) | \Sigma) = x_{diff} t_d / t^*$ and the expression for $x_1$ becomes the standard one for a jump totally evolving in an homogeneous medium, i.e $x_1 = x_0 + \Lambda^{-1}(R_x^1 | \Sigma) + vt_1$.

If on the contrary $x_{diff}$ is negative $(R_1^x < 0.5)$ (cf. Fig.4, bottom), then $x_{diff} < x_{diff} t_d / t^* \leq 0$ and correspondingly $\Lambda(x_{diff} t_d / t^* | \Sigma_1) > R_1^x$. The additional diffusive component, as given by the difference between the abscissas of the points P$_1$ and P$_2$ on $\Lambda(x | \Sigma_2)$, is directed backwards, against the advection velocity field. To this quantity we must add the contribution $v(t_1 - t_d)$ of the advection, so that the jump ends at the same value given by Eq. (10). However, now the term in square parentheses is negative and it might happen that $x_1 < x_d$. This condition would mathematically mean that the walker, as soon as arrived at $x_d$ driven by $v$, is driven back to layer 1 by the diffusive component in $L_2$ instead of proceeding further. Here, the advection velocity pushes it again towards layer 2 and these back and forth "oscillations" continue up to time $t_1$, with the walker always stuck in $x_d$. To take into account this possibility, Eq. (10) is modified as follows:

$$x_1 = x_d + \max\{0, [\Lambda^{-1}(R_1^x | \Sigma_2) - \Lambda^{-1}(\Lambda(x_{diff} t_d / t^* | \Sigma_1) | \Sigma_2)] + v(t_1 - t_d)\} \tag{11}$$

We now consider a jump starting at $x_0$ in $L_2$. As before, two uniform random numbers $R_2^x$ and $R_2^t$ are sampled and the jump duration and the diffusive contribution to the length are then $t^* = W^{-1}(R_2^t | \tau_2, \alpha_2)$ and $x_{diff} = \Lambda^{-1}(R_2^x | \Sigma_2)$, respectively. Note that in this case the walker is driven forwardly by the advection velocity, so that it may hope to go back to $L_1$ only if $x_{diff}$ is negative, i.e. if $R_2^x < 0.5$. The total jump length would be $x_{diff} + vt^*$ if the new location were still in $L_2$. However, $x_{diff}$ may be so negative that the walker can cross the discontinuity located at $x_d$ after having utilized the portions $t_d = \dfrac{x_d - x_0}{x_{diff} + vt^*} t^*$ and $x_{diff} \dfrac{t_d}{t^*}$ of the assigned $t^*$ and $x_{diff}$ and the portions $W(t_d | \tau_2, \alpha_2)$ and $\Lambda(x_{diff} t_d / t^* | \Sigma_2)$ of the assigned $R_2^x$ and $R_2^t$.

The analysis is similar to that for $x_0 \in L_1$ and $x_{diff} < 0$ and the result is given by Eq. (10) with reversed $\Sigma_1$ and $\Sigma_2$, but now we must discard the possibility that the jump could end within $L_2$, that is $x_1 > x_d$, so that

$$x_1 = x_d + \min\{0, [\Lambda^{-1}(R_2^x | \Sigma_1) - \Lambda^{-1}(\Lambda(x_{diff} t_d / t^* | \Sigma_2) | \Sigma_1)] + v(t_1 - t_d)\} \tag{12}$$

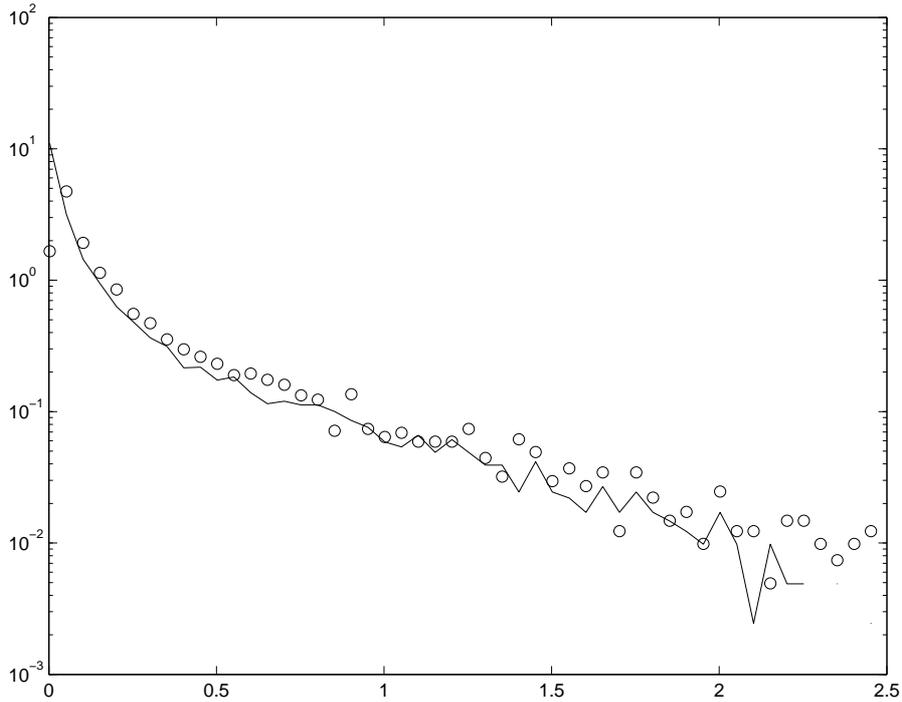

*Figure 5. Distributions of jump times across a discontinuity obtained by utilizing a single random number (solid curve) or two random numbers (circles).*

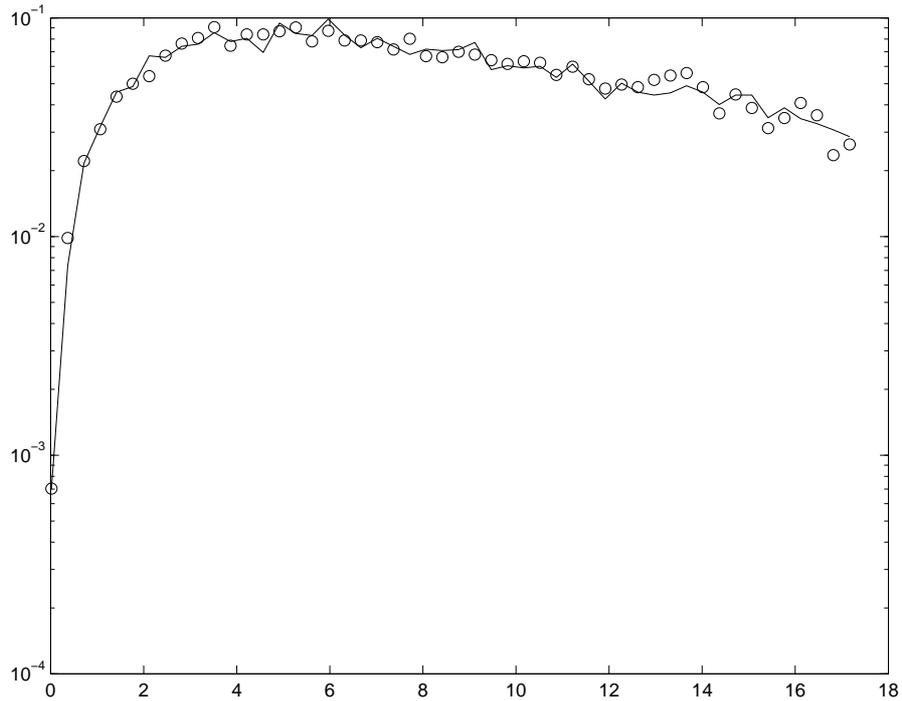

*Figure 6. Distributions of jump lengths across a discontinuity obtained by utilizing a single random number (solid curve) or two random numbers (circles).*

Until now, each jump has been simulated making use of one random $R_1^t$ for the time duration and one random $R_1^x$ for the total length. However, as for the simple exponential, we might also simulate the whole jump by resorting to two pairs of random numbers: the portion of the travel in $L_1$ with $R_1^t$ and $R_1^x$ as before and the portion in $L_2$ with two new random numbers $R_t^*$ and $R_x^*$. Figures 5 and 6 show the results of the two approaches.

Differently from the exponential case, now we must take into account that the problem is semi-Markovian, so that, once the walker has reached the interface between $L_1$ and $L_2$,

*i)* the probability of the jump further lasting from $t>t_d$ to $t+dt$ depends on the sojourn time $t_d$ already elapsed and *ii)* the probability of further travelling from $x > x_d$ to $x+dx$ depends on the diffusive component $x_{diff} \, t_d / t^*$ of the travel already effectuated: correspondingly, the new random numbers must be sampled no more within $[0,1)$, but in suitably restricted intervals.

The time to be spent in $L_2$ is simulated by sampling a uniform number $R_t^*$ in the interval which goes from the value utilized in $L_1$ up to unity, i.e. $[W(t_d | \tau_1, \alpha_1), 1)$, and the whole duration is given by Eq. (9), in which $R_t^*$ takes the place of $R_1^t$. Concerning the space to be travelled in $L_2$, the interval in which to sample the new random $R_x^*$ starts from $\Lambda(x_{diff} t_d / t^* | \Sigma_1)$, the value utilized in $L_1$ (the ordinate of point $P_2$ in Fig. 4), and it must contain $R_x^1$: thus, when $x_{diff}$ is positive, the interval is $[\Lambda(x_{diff} t_d / t^* | \Sigma_1), 1)$, while in the opposite case the interval is $[0, \Lambda(x_{diff} t_d / t^* | \Sigma_1))$. The whole jump length is given by Eqs. (11) and (12) in which $R_x^*$ takes the place of $R_1^x$.

Before leaving the Galilei invariant case, we remark that, within the classical CTRW model, the motion of a walker in a homogeneous medium is evaluated by first analytically solving the problem with $v=0$ (with the walker trapped at the start of each jump for the whole jump time) and by successively shifting the solution thereby obtained. The final concentration is then $P(x,t)=P_{v=0}(x-vt,t)$. Clearly, the shifting operation implies a linear motion of the walker along the jump.

Next, we consider the Galilei variant case: at each jump, the effect of the advection is modelled by adding to the diffusive contribution a constant term $\mu$. The situation is now much simpler, because the space and time variables are independent of each other and because the walker cannot be stuck at the discontinuity. Indeed, the presence of the bias $\mu$ may be taken into account by sampling the jump lengths from shifted Gaussian distributions $\Lambda(x | \mu, \Sigma_i)$. Correspondingly, the two portions of a jump across a discontinuity cannot have opposite signs. The formulae given for the Galilei invariant case still hold true, provided that $v$ is set to zero and all the $\Lambda$'s cdf are assumed with a mean $\mu$ instead of zero. Then, the *min* and *max* operators in eqs. (11) and (12) play no role and must be taken out.

Before ending this section, we would like to present a final remark: we have numerically verified that – in absence of a discontinuity – in the Galilei variant case the assumption of a linear motion for the walker leads to concentration profiles which coincide with those one would have obtained by adopting the classical CTRW model, i.e. a motion constituted by waiting times followed by sudden jumps at infinite velocity. This equivalence has been tested under the assumption $x \gg \Sigma + \mu$, i.e. in the "diffusion limit" approximation.

**5. Monte Carlo estimates of flux and concentration**

Resident concentration is physically defined as the mass of tracer particles per unit volume of flowing substance contained in an elementary volume of the medium at a given time. Flux is defined as the mass of tracer particles passing through a given cross section during an elementary time interval [27,17]. It should be remarked that flux is the most commonly measured variable in experiments determining breakthrough curves at the end of a test column where tracer particles are free to diffuse. In a Monte Carlo simulation, the phase space $\{t, \vec{x}\}$ is discretized in cells where weights are accumulated. When a walker moves across a sequence of cells, the contribution of its trajectory to the mean flux in a given cell at a given (discrete) time is obtained by accumulating the length of the travel spanned within the cell at that time. Analogously, the contribution of the trajectory to the mean resident concentration in a given cell at a given (discrete) position is obtained by accumulating the time length spent within the cell at that position [28]. In our case of one-dimensional transport, the phase space is discretized with a rectangular grid composed of identical cells of area $dx\,dt$. Thus, the contribution to the flux in each cell is constant and equal to $dx$ and similarly the contribution to the concentration is constant and equal to $dt$.

With reference to Figure 7, the algorithm for flux collection along each jump starting at a point *A* in the phase space and ending in *B* proceeds as follows. Starting from *A*, consider in succession the upper (if *B* is above *A* with respect to the *x*-axis) or lower (if *B* is below *A*) extremes of the consecutive $dx$ spanned by the trajectory (i.e. the straight line connecting *A* to *B*). For each extreme $x_q$ of cell $dx_q$ determine the corresponding point *q* on the segment *A-B* and the time channel $dt_q$ to which *q* pertains. A weight $dx$ is finally added to the $\phi(t,x)$ matrix at position $\{dt_q, dx_q\}$.

The algorithm for concentration collection is symmetric with respect to the previous one. With reference to Figure 8, along each jump starting at a point *A* in the phase space and ending in *B* it proceeds as follows. Starting from *A*, consider in succession the right extremes of the consecutive $dt$ spanned by the trajectory. For each extreme $t_q$ of cell $dt_q$ determine the corresponding point *q* on the segment *A-B* and the spatial channel $dx_q$ to which *q* pertains. A weight $dt$ is finally added to the $P(t,x)$ matrix at position $\{dt_q, dx_q\}$.

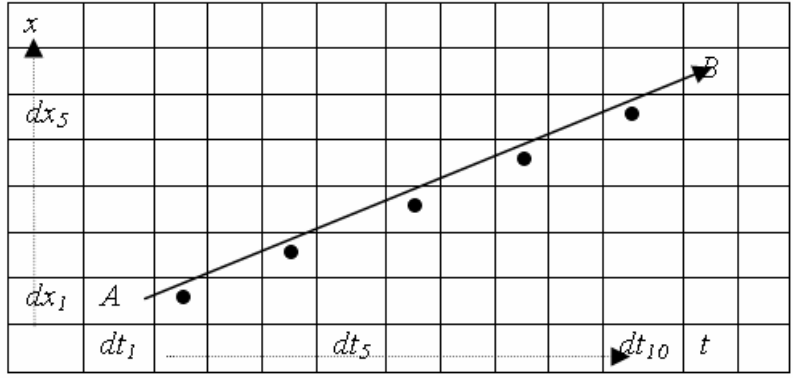

*Figure 7. An example of collection of contributions to mean flux.*

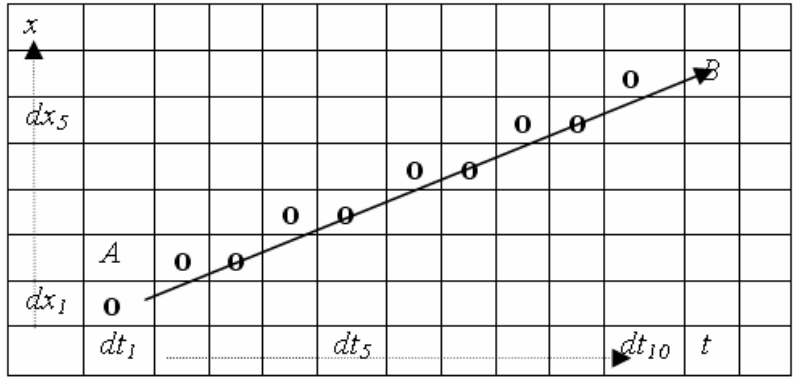

*Figure 8. An example of collection of contributions to mean concentration.*
*Label A refers to the cell ($dt_1$, $dx_1$), as in Figure 7.*

## 6. Results

In this Section we present two examples of anomalous transport across a discontinuity, one referring to the Galilei invariant (Figure 9) and the other to the Galilei variant scheme (Figure 10). No constraints on the chosen pdf's parameters have been imposed, so that we will not discuss the physical relevance of the present results. Both concentration $P(x,t)$ and flux $\phi(x,t)$ spatial profiles at some successive times are shown. Flux is always continuous at the interface, as required by mass conservation principle: the same number of particles crossing a surface must be found on the other side, as no adsorption phenomena take place during the crossing. However, in general the slope of the flux will vary while crossing the discontinuity. On the other hand, for the pdf's parameters here adopted, resident concentration profiles are shown to present neat discontinuities at the interface. A similar situation has been reported in standard advection-dispersion [29-31] and in simplified anomalous advection-dispersion models [32]. The discontinuity in resident concentration necessarily implies the presence of abruptly varying physical properties, as it may occur in subsurface transport in case of a sudden separation between different geological materials [33]. However, from an analytical point of view (with reference to the standard advection-dispersion equation), resident concentrations at the interface between two different media may be either continuous or discontinuous according to the choice of the boundary conditions and this matter has been the subject of considerable discussions [34-43]. Continuity is obviously recovered if resident concentrations are required to be equal at the interface [33,41-43], while discontinuity arises if the concentrations at the interface are left free, while the gradient of the resident concentration of the left layer (with respect to flow direction) is imposed to vanish at infinity [29-36]. Correspondingly, also the analytical flux shapes are different and in real cases, in order to have a suitable model, one selects the proper boundary condition which best accounts for the measured flux shape (which is usually the experimentally accessible variable) [37]. The Monte Carlo simulation parameters we adopted in this paper are consistent with the assumption of a concentration gradient vanishing at infinity proposed by e.g. [34,36-37]: this situation has been reported to occur e.g. at the interface between an experimental column and its exit reservoir (considered as left and right layer, respectively [37-39]) in contaminant tracers transport phenomena. It should be remarked that the discontinuity in resident concentrations is relevant only when the advection contribution is small with respect to the dispersion (i.e. at low Péclet numbers [37]) and is completely negligible otherwise.

In Figure 9 we consider the travels of particles starting in $x=0$ at $t=0$ and show the spatial flux and concentration profiles across a discontinuity occurring at $x_d=30$ at three successive times, for a Galilei invariant scheme with velocity $v=5$. The simulation parameters are the following: $\Sigma_1=3$, $\Sigma_2=2$, $\alpha_1=0.9$, $\alpha_2=0.4$, $\tau_1=\tau_2=0.01$. One million particles have been followed up to a final time $t_{max}=3>>\tau$. With respect to the resident concentration, the discontinuity seems to act as a source for the right layer and the velocity translates both the original peak located at $x=0$ and the new peak located at $x_d$, thus enhancing the smoothing of the concentration peak in correspondence of the discontinuity. The change in slope in the flux profile in correspondence of the concentration discontinuity is neatly evident.

In Figure 10 we show the spatial flux and concentration profiles across a discontinuity occurring at $x_d=30$ at three successive times, for a Galilei variant scheme with a bias $\mu=0.5$. The simulation parameters are the same as in the previous case. Again, the discontinuity seems to act as a source of resident concentration. In the Galilei variant scheme, however, the peak located at the source drastically decreases with time but it does not move from $x=0$, while the concentration profile stretches in the bias direction: this effect is particularly evident in the portion of the medium at the right of the discontinuity, where subdiffusion is neatly dominant. A prominent peak accumulates particles which are then slowly dispersed by the combined effect of the bias and of the diffusion contribution. The change in slope of the flux profile is less evident than in the Galilei invariant scheme.

We would like to remark that the results here obtained, and in particular the presence of a macroscopic discontinuity in resident concentration at the interface, depend on the fact that within the Monte Carlo approach the distributions parameters may be arbitrarily chosen. Imposing constraints on these parameters leads to families of solutions in which the observed discontinuity in resident concentration varies and may even disappear. This situation is analogous to the one of the analytical approach (for the standard ADE), where different boundary conditions give rise to either discontinuous or continuous solutions. A work aimed at establishing the connection between MC parameters, boundary conditions and related physical implications is in progress.

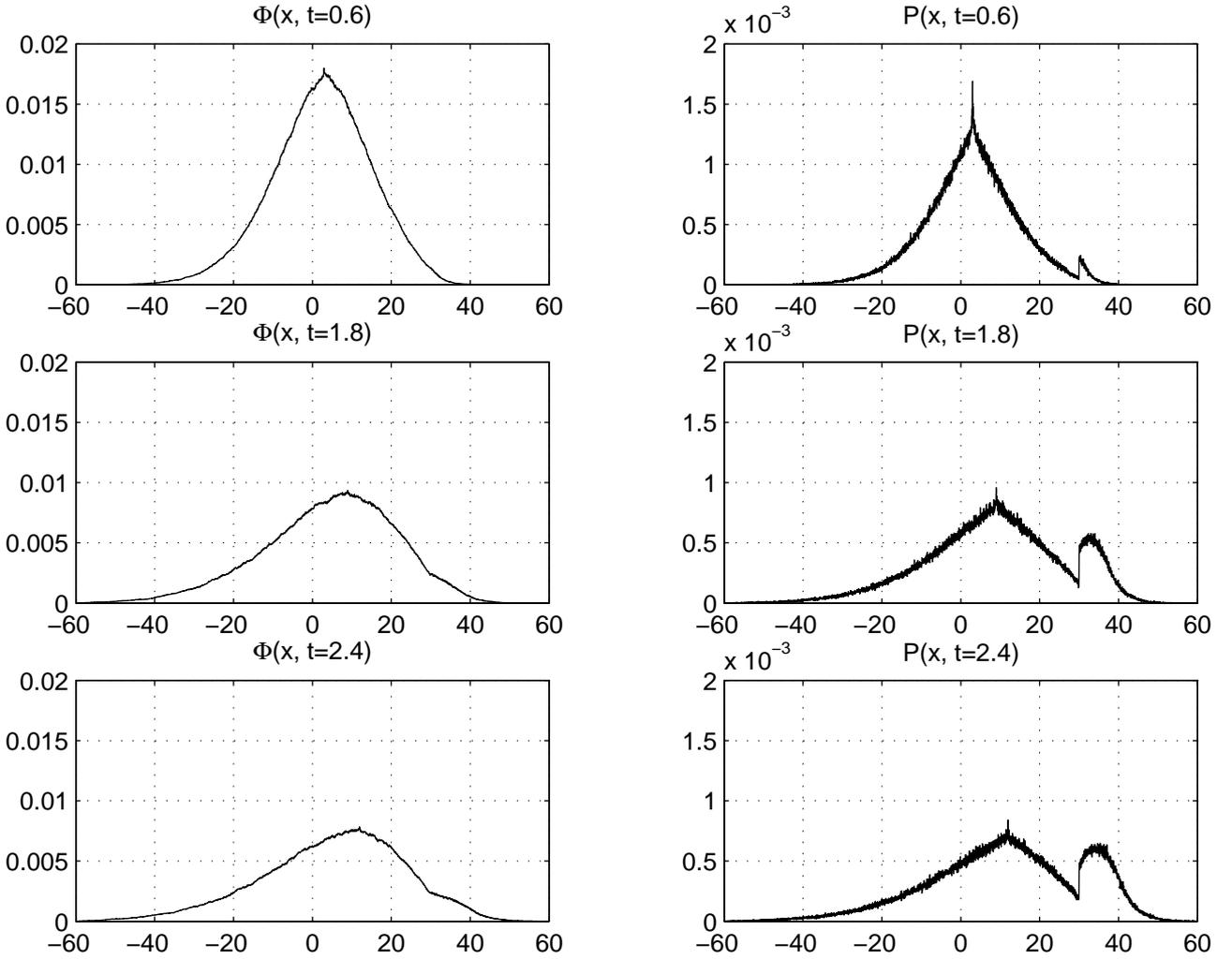

*Figure 9. Galilei invariant scheme: spatial profiles of flux and concentration across a discontinuity at three given times.*

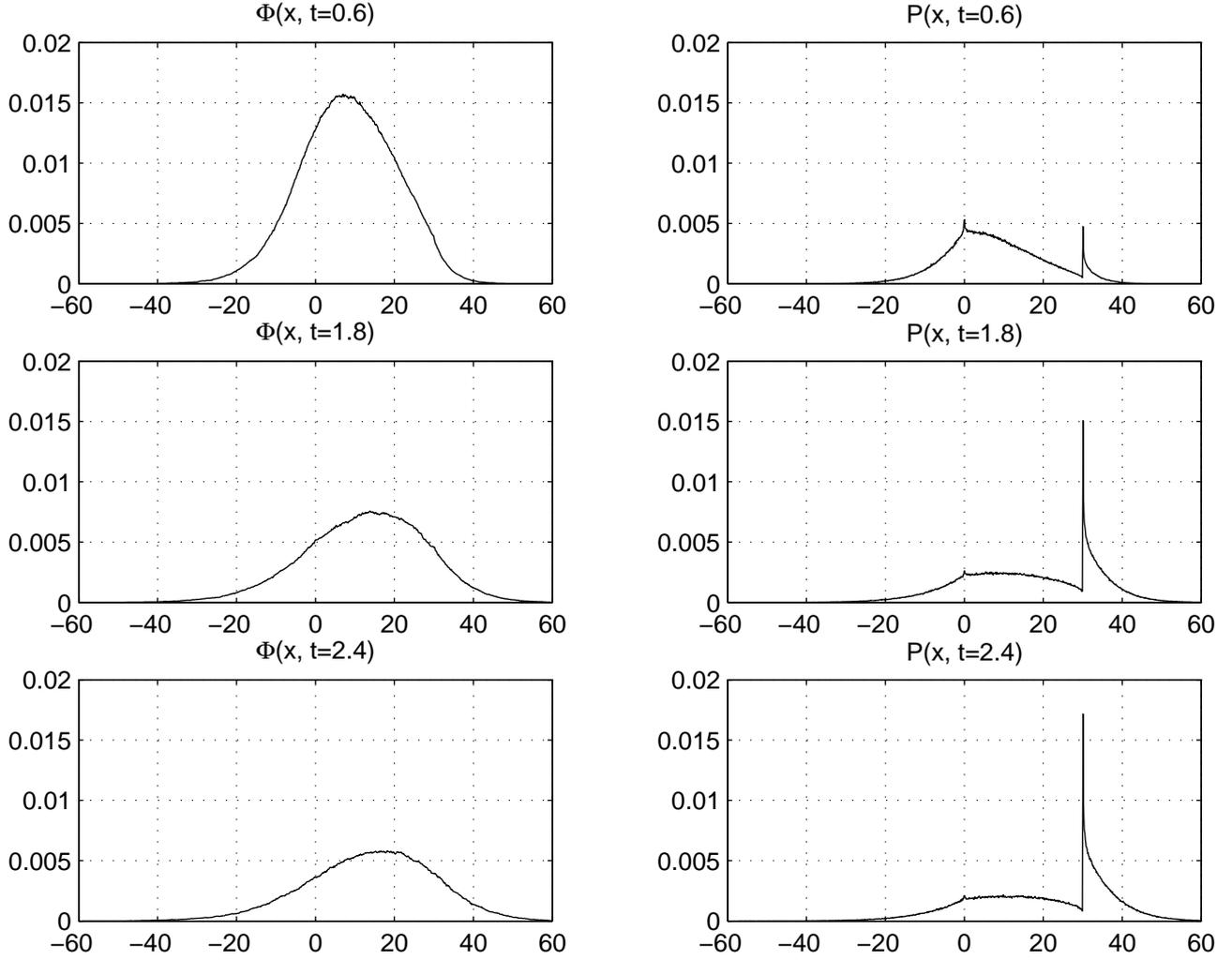

*Figure 10. Galilei variant scheme: spatial profiles of flux and concentration across a discontinuity at three given times.*

### 7. Conclusions

Experimental evidences of anomalous transport, occurring in quite different contexts, have been successfully interpreted in the framework of the Continuous Time Random Walk, in which an ensemble of particles (walkers) is supposed to move stochastically in a medium according to spatial and temporal coordinates sampled from given distributions. At each jump, the walker is supposed to stay at rest at the starting point for the whole jump time and then suddenly jump (at infinite velocity) to the new location. As long as the properties of the medium where the transport process takes place are stationary, i.e. the parameters of the distributions from which space and time coordinates of the walker are sampled are constant, known analytical (approximated fractional advection-diffusion schemes) or numerical solutions exist. In this paper, within the Monte Carlo simulation approach to CTRW, we have investigated the transport in a non-homogeneous medium constituted by two regions with different physical properties, in presence of an additional advection, which encourages transport in the forward direction. The velocity at which the walker performs each jump is thus the result of the diffusive contribution and of the advection field: advection has been modeled within two schemes, namely the Galilei invariant and the Galilei variant schemes, each pertaining to different physical situations. Differently from the standard CTRW approach, throughout this paper it is assumed that particles move linearly at constant speed (variable from jump to jump) between successive interactions. For such a situation – to the authors' best knowledge – no analytical solution is readily available. The general approach which allows to sample jumps across a boundary has been at first illustrated by means of an example involving Markovian particles. Successively, this method has been extended to a semi-Markovian process and applied to the subdiffusive transport in both a Galilei invariant and a Galilei variant advection scheme. After having briefly illustrated the Monte Carlo collection technique which allows to calculate the mean particle concentration and flux, two examples of concentration and flux profiles have been presented and discussed. We have shown that the flux distribution presents a discontinuity in its slope but preserves continuity, as required on physical bases (mass conservation). On the contrary, for the choice of the distribution parameters here adopted, the presence of the discontinuity has the macroscopic effect of introducing a neat discontinuity in the particle concentration profile at the interface. The obtained results, and in particular the presence of a macroscopic discontinuity in resident concentration at the interface, depend on the fact that within the Monte Carlo approach the distributions parameters may be arbitrarily chosen: imposing constraints on these parameters leads to families of solutions in which the observed discontinuity in resident concentration varies and may even disappear. This situation is analogous to the one of the analytical approach (for the standard ADE), where different boundary conditions give rise to either discontinuous or continuous solutions. Monte Carlo simulation has proven to be a very effective way to investigate the statistical properties of anomalous transport across an interface: the results here obtained could be generalized to more involved and more realistic situations, e.g. with distribution parameters varying smoothly in space and in time.

**Acknowledgements**

This work was supported by the Italian Ministry of Education, University and Research (MIUR), through a project titled *Transport of toxic and/or radioactive contaminants through natural and artificial porous media: models and experiments*.

**Appendix. Some considerations on the advection field**

We would like to present some considerations on the Galilei invariant and Galilei variant advection schemes introduced in Section 4. The Galilei invariant approach to model advection assumes that the tracer particle is dragged by an homogeneous velocity field, namely $v$. Under this assumption, the particle concentration at a time $t$ and at the position $x$-$vt$ is the same which would have been measured at time $t$ and at location $x$ if no advection field had been present: this consideration justifies the name "Galilei invariant". A uniform velocity field is found to occur e.g. in the motion of tracer particles in a flow field when the flowing substance is itself the cause of (anomalous) diffusion, as in polymer solutions [1]. On the contrary, in the Galilei variant scheme the walker is assumed to perform biased jumps with a mean value $\mu$ which may be thought of as the product of a constant speed by a constant characteristic advection time. This scheme does not lead to the Galilei invariance principle previously described and has been successfully invoked to explain evidences of anomalous transport in porous media, where particles get trapped before being released into the velocity field again [13-20,1-3].

Let us now consider a generic jump and assume that $x$ and $t$ are sampled from the same distributions for both the Galilei invariant and Galilei variant schemes, i.e. a zero-mean Gaussian and a power law pdf's, respectively. For sake of simplicity, we assume that no discontinuity occurs. Thus, in the first case the jump will be performed at a velocity $v_{inv} = \frac{x+vt}{t} = \frac{x}{t} + v$, while in the second case $v_{var} = \frac{x+\mu}{t} = \frac{x}{t} + \frac{\mu}{t}$. In this way we have decomposed the velocity in a common diffusive contribution, $\frac{x}{t}$, with zero mean, plus an advective bias, which is constant for the invariant case and stochastic for the variant case. In order to compare the two cases, we choose $\mu$ so that $\left\langle \frac{\mu}{t} \right\rangle = v$. While the two distributions of $v_{var}$ and $v_{inv}$ are visually similar (Figure 11), those of the biases $\frac{\mu}{t}$ and $v$ are quite different (Figure 12). These diversities are neatly reflected in the shape of the particle concentration for the two advection schemes, as shown in the following Figure 13. In case of Galilei invariant motion, the maximum of the concentration moves in the direction of the bias so that $\langle x(t) \rangle = vt$ [1]. In case of Galilei variant transport, on the contrary, the concentration peak stays at the origin, while the concentration itself stretches towards the bias direction: it can be shown that $\langle x(t) \rangle \approx t^\alpha$ [20]. There are experimental evidences that the Galilei variant model applies to the contaminant transport in porous or fractured media [13-20]. When the waiting times are sampled from an exponential pdf with mean $\tau$, the Galilei variant and invariant schemes become indistinguishable if we set $v = \frac{\mu}{\tau}$.

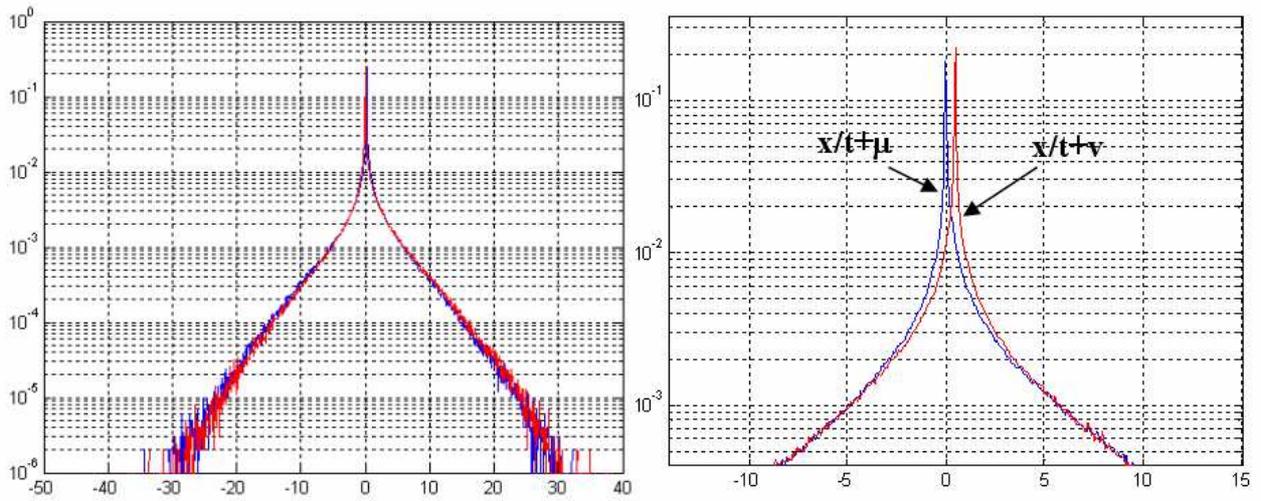

*Figure 11. Left: distribution of $v_{var}$ and $v_{inv}$. ($\mu = 0.1$, $v = \left\langle \frac{\mu}{t} \right\rangle \cong 0.23$). Right: zoom.*

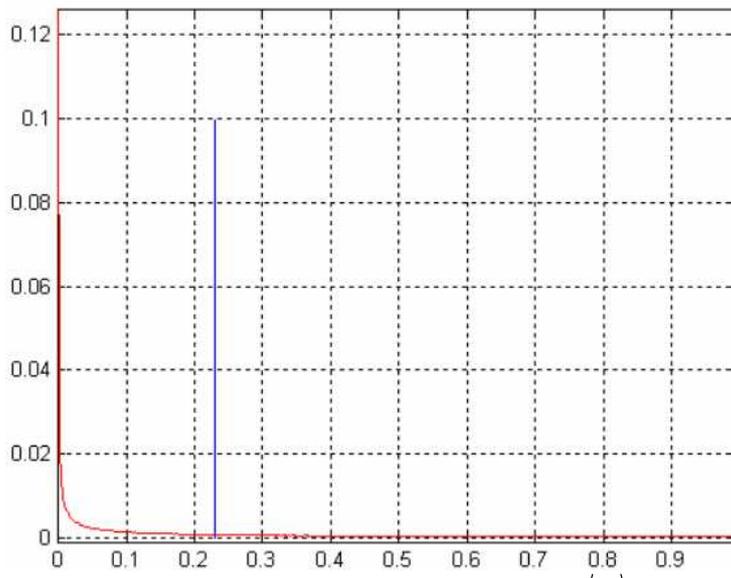

*Figure 12. Distribution of $\frac{\mu}{t}$ and v. ($\mu = 0.1$, $v = \left\langle \frac{\mu}{t} \right\rangle \cong 0.23$)*

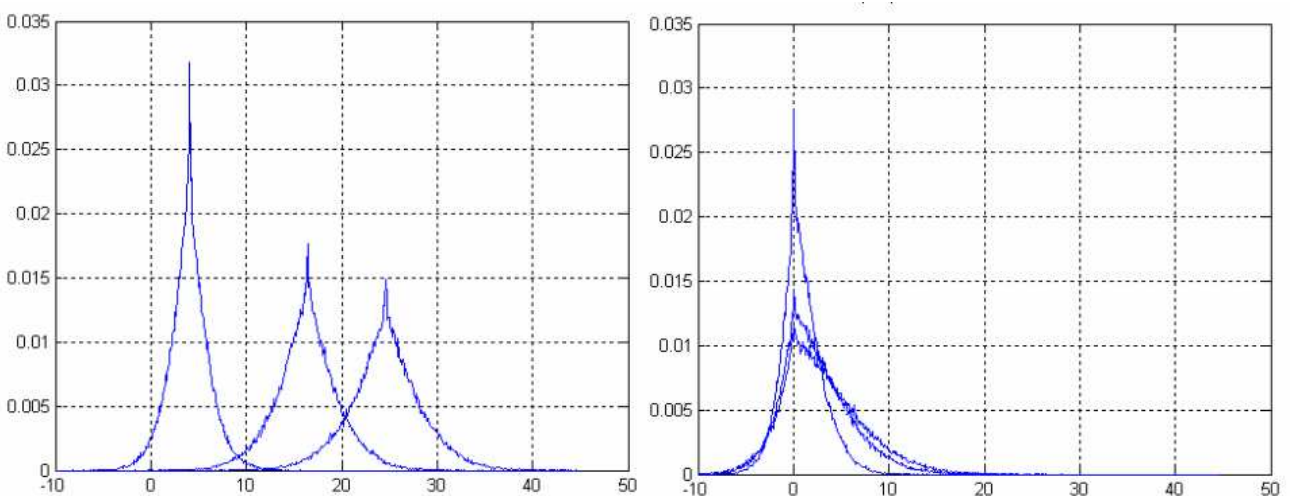

*Figure 13. Left: evolution of concentration (at three successive times) in a Galilei invariant scheme. Right: evolution of concentration (at the same times as at the left side figure) in a Galilei variant scheme.*